\documentclass[letters,usenatbib]{mnras}

\usepackage[T1]{fontenc}
\usepackage{ae,aecompl}

\usepackage{graphicx}	
\usepackage{amsmath}	
\usepackage{amssymb}	

\def\lya{{\rm\,Ly$\alpha$}}
\def\ha{{\rm\,H$\alpha$}}
\def\oii{{\rm\,[O{\sc ii}]}}
\def\nii{{\rm\,[N{\sc ii}]}}

\def\hi{{\rm\,H{\sc i}}}
\def\msun{{\rm M}$_{\odot}$}

\title[Ly$\alpha$ depletion in the protocluster core]{Direct evidence for Ly$\boldsymbol\alpha$ depletion in the protocluster core}

\author[R. Shimakawa et al.]{
Rhythm Shimakawa,$^{1,2}$\thanks{rhythm.shimakawa@nao.ac.jp} 
Tadayuki Kodama,$^{1,2}$\thanks{t.kodama@nao.ac.jp}
Masao Hayashi,$^{2}$
Ichi Tanaka,$^{3}$
\newauthor
Yuichi Matsuda,$^{1,2}$
Nobunari Kashikawa,$^{1,2}$
Takatoshi Shibuya,$^{4}$
Ken-ichi Tadaki,$^{5}$
\newauthor
Yusei Koyama,$^{1,3}$
Tomoko L. Suzuki$^{1,2}$
and Moegi Yamamoto$^{1}$
\\
$^{1}$Department of Astronomical Science, SOKENDAI, Osawa, Mitaka, Tokyo 181-8588, Japan\\
$^{2}$National Astronomical Observatory of Japan, Osawa, Mitaka, Tokyo 181-8588, Japan\\
$^{3}$Subaru Telescope, National Astronomical Observatory of Japan, 650 North A'ohoku Place, Hilo, HI 96720, USA\\
$^{4}$Institute for Cosmic Ray Research, The University of Tokyo, 5-1-5 Kashiwanoha, Kashiwa, Chiba 277-8582, Japan\\
$^{5}$Max-Planck-Institut f\"{u}r Extraterrestrische Physik, Giessenbachstrasse, D-85748 Garching Germany
}

\date{Accepted 2017 January 31. Received 2017 January 18; in original form 2016 December 22}

\pubyear{2017}

\begin{document}
\label{firstpage}
\pagerange{\pageref{firstpage}--\pageref{lastpage}}
\maketitle

\begin{abstract}
We have carried out a panoramic Ly$\alpha$ narrowband imaging with Suprime-Cam on 
Subaru towards the known protocluster USS1558--003 at $z=2.53$. Our previous 
narrowband imaging at near-infrared has identified multiple dense groups of 
H$\alpha$ emitters (HAEs) within the protocluster. We have now identified the 
large-scale structures across a $\sim$50 comoving Mpc scale traced by Ly$\alpha$ 
emitters (LAEs) in which the protocluster traced by the HAEs is embedded. On a 
smaller scale, however, there are remarkably few LAEs in the regions of HAE 
overdensities. Moreover, the stacking analyses of the images show that HAEs in 
higher-density regions show systematically lower escape fractions of Ly$\alpha$ 
photons than those of HAEs in lower-density regions. These phenomena may be driven 
by the extra depletion of Ly$\alpha$ emission lines along our line of sight by 
more intervening cold circumgalactic/intergalactic medium and/or dust existing in 
the dense core. We also caution that all the past high-$z$ protocluster surveys 
using LAEs as the tracers would have largely missed galaxies in the very dense 
cores of the protoclusters where we would expect to see any early environmental 
effects.

\end{abstract}




\section{Introduction}

High-$z$ galaxy protoclusters \citep{Sunyaev:1972} are ideal test beds where we 
can understand how cluster galaxies form and grow during the course of cosmic 
mass-assembly history and the build-up of large-scale structures (LSSs) in the 
Universe \citep{White:1991,Cole:2000}. They directly inform us of what is 
occurring in the early phase of cluster formation and galaxy formation therein, 
which then tells us what the physical mechanisms are that lead to the galaxy 
diversity depending on the environment seen in the local Universe 
\citep{Dressler:1980,Cappellari:2011}. 

The observational limitation due to the Earth's atmosphere has created a gulf 
between high-$z$ galaxy surveys at $z<2.6$ and those at $z>2.6$. At redshifts 
greater than 2.6, bright \ha$\lambda6565$ emission line is no longer observable 
from ground-based telescopes, and the \lya$\lambda1216$ line is the most commonly 
used spectral feature of star-forming galaxies that can be captured by optical 
instruments. This technique has been widely used to identify galaxies at high 
redshifts both in the general fields and in overdense regions such as 
protoclusters \citep{Ouchi:2003,Venemans:2007}. However, we know that only a small 
fraction of star-forming galaxies show detectable \lya\ emission lines 
\citep{Hayes:2010,Matthee:2016,Hathi:2016}. Furthermore, the environmental 
dependence of \lya\ emitters (LAEs) has largely been unexplored yet. Therefore, 
understanding the dependence of physical properties and the selection effects of 
LAEs across various environments is strongly desired.

In this respect, the dual emitter surveys of HAEs and LAEs for the known 
protoclusters at $z=$ 2.1--2.6 can play a key role in testing the \lya\ selection 
effect for high-$z$ protocluster search. This {\it letter} studies \lya\ 
emissivities of \ha-emitting galaxies in a known dense protocluster, USS~1558--003 
at $z=2.53$ 
($\alpha_\mathrm{J2000}=$ 16$^\mathrm{h}$01$^\mathrm{m}$17$^\mathrm{s}$ 
$\delta_\mathrm{J2000}=$ $-$00$^\mathrm{d}$28$^\mathrm{m}$47$^\mathrm{s}$, 
hereafter USS~1558) discovered by \citet{Kajisawa:2006,Kodama:2007,Hayashi:2012} 
with MOIRCS on the Subaru telescope. Our recent deep \ha\ narrowband survey of 
this region succeeded in detecting more protocluster members amounting to 100 HAEs 
in total \citep{Hayashi:2016}, allowing us to characterise sub-structures such as 
clumps associated to USS~1558. Given such a unique laboratory of overdense 
environment at $z=2.5$, we have manufactured a dedicated narrowband filter for 
this specific target and installed it on the Suprime-Cam so that we can also 
target LAEs associated with this region. Also, the wider field of view of 
Suprime-Cam (32'$\times$27') compared to MOIRCS (7'$\times$4') enables us to map 
LSSs in and around USS~1558.

We assume the cosmological parameters of $\Omega_M=0.3$, $\Omega_\Lambda=0.7$ and 
$h=0.7$ and employ a \citet{Chabrier:2003} stellar initial mass function, and the 
AB magnitude system \citep{Oke:1983} are used throughout the {\it Letter}. 
Galactic extinctions at NB428 and B-band are assumed to be 0.55 and 0.57 mag, 
respectively \citep{Schlegel:1998,Fitzpatrick:1999,Schlafly:2011}\footnotemark[1].

\footnotetext[1]{\url{http://irsa.ipac.caltech.edu/applications/DUST/}}

\section{Observation and data analyses}

\begin{figure}
	\centering
	\includegraphics[width=0.95\columnwidth]{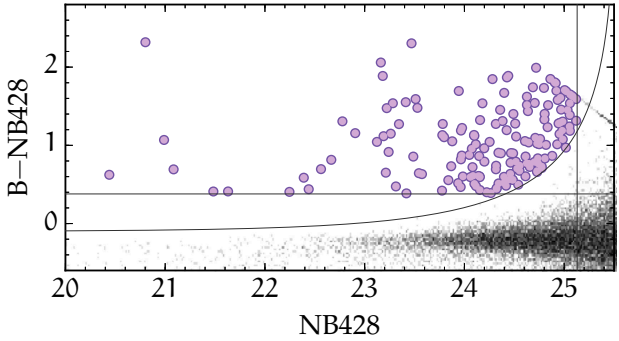}
    \caption{The colour--magnitude diagram. The black shows the NB-detected 
    sources. The purple circles indicate narrow-band emitters that meet our 
    criteria, namely (1) the line EW is greater than $>15$ \AA\ in the 
    rest-frame (the horizontal line), (2) line flux excess is larger 
    than $3\sigma$ (the black curve), and the detection at NB428 is more 
    than $5\sigma$ levels (the vertical line).}
    \label{fig1}
\end{figure}

We performed \lya\ line imaging of USS~1558 at $z=2.53$ with the Subaru Prime 
Focus Camera (Suprime-Cam; \citealt{Miyazaki:2002}) on the Subaru telescope. We 
used the custom-made narrowband filter NB428 that has the central wavelength of 
4297 \AA\ and FWHM of 84 \AA. This is designed so that the filter FWHM neatly 
captures the \lya\ lines at $z=2.53\pm0.03$ (emission or absorption) from HAEs at 
$z=2.52\pm0.02$ associated with the USS~1558 protocluster selected by 
the narrowband filter, NB2315, installed on MOIRCS/Subaru (see 
\citealt{Shimakawa:2016})\footnotemark[2]. The combined analysis of a resonant 
\lya\ line by NB428 
and a non-resonant \ha\ line by NB2315 enables us to make the first systematic 
comparison of spatial distributions between LAEs and HAEs, and to investigate the 
environmental dependence of the \lya\ photon escape fractions within the 
protocluster.

\footnotetext[2]{The bandpass of NB428 for \lya\ does not 
perfectly match to that of NB2315 for \ha. This leads to 10 \% \lya\ flux loss on 
average for HAEs which is considered in our stacking analyses (\S4). Also, 
the NB2315 bandpass shifts blueward towards the edge of MORICS field of view, 
however, this effect can be negligible for our HAE samples according to the 
past follow-up spectroscopy \citep{Shimakawa:2014}. Furthermore, 
NB428 covers up 
to higher redshifts for \lya\ by $\gtrsim1500$ km~s$^{-1}$ with respect to that of 
NB2315 for \ha, and thus a possible redward velocity offset of \lya\ relative to 
\ha\ \citep{Shapley:2003} would be negligible.}

The observation was executed on June 10 in 2015 under a photometric condition 
but with a relatively bad seeing (FWHM=0.8--1.4 arcsec). The science frames with 
seeing sizes worse than 1.3 arcsec were trashed and we used 19 frames of 700 sec 
exposures each, amounting to 3.7 hrs of net integration time in total. The data 
were reduced in exactly the same way as in \citet{Shimakawa:2016} based on a data 
reduction package for the Suprime-Cam, {\sc sdfred} (ver.2; 
\citealt{Yagi:2002,Ouchi:2004}). The pipeline includes the standard procedures. 
The sky subtraction was conducted with the mesh size of 13 arcsec. We additionally 
implemented a cosmic ray reduction using the algorithm, L.A.Cosmic 
\citep{Dokkum:2011}. The final combined image has a seeing size of FWHM = 1.24 
arcsec, and the limiting magnitude of 25.13 mag at 5 sigma with 2.5 arcsec 
aperture diameter reckoning with the galactic extinction. 

Combining the reduced NB428 data with the existing counterpart B-band image 
(B$_\mathrm{5\sigma}=25.72$ mag with 2.5 arcsec aperture diameter) provided by 
\citet{Hayashi:2016}, we select LAE candidates in the same manner as in 
\citet{Shimakawa:2016}. Here, the seeing FWHM of B-band is tuned to that of the 
NB428 image. The object photometry is performed by SExtractor (ver.2.19.5; 
\citealt{Bertin:1996}). Photometric measurements are done in the double image mode 
using the narrowband image for source detections, and we employed aperture 
photometries with 2.5 arcsec diameter. This work imposes the selection criteria as 
follows; (1) narrowband flux excess is greater than $3\sigma$ with respect to the 
photometric error, (2) \lya\ equivalent width is higher than 
EW$_\mathrm{Ly\alpha}$ = 15 \AA\ in the rest frame, and (3) 
$NB_\mathrm{5\sigma}<25.13$ mag. The correction factor of the colour term (i.e. 
zero point of B$-$NB) is assumed to be $-0.1$, which is tailored to that in our 
past study of another field using the same NB428 filter \citep{Shimakawa:2016}. 
$2\sigma$ limiting magnitude are assumed for the sources with B-band detection 
levels lower than $2\sigma$. 

As a result, in total 162 objects satisfied our selection criteria 
(Fig.~\ref{fig1}). This {\it Letter} employs these samples as LAE candidates at 
$z=2.5$. However, it is noted that a considerable number of foreground 
contaminations such as \oii$\lambda\lambda3727,3730$, 
C{\sc iv}$\lambda\lambda1548,1551$ would be contained even though we are targeting 
the region that hosts the known protocluster (c.f.\ $\sim60$ \% in the random 
field according to \citealt{Sobral:2016}). The currently available photometric 
data that cover the entire field are only B, r, and z-band photometries with 
Suprime-Cam, which are not sufficient to cleanly decontaminate our LAE samples. 
This caution should be kept in mind when our LAE samples are discussed.

\section{Results} 

\begin{figure*}
	\includegraphics[width=1.0\textwidth]{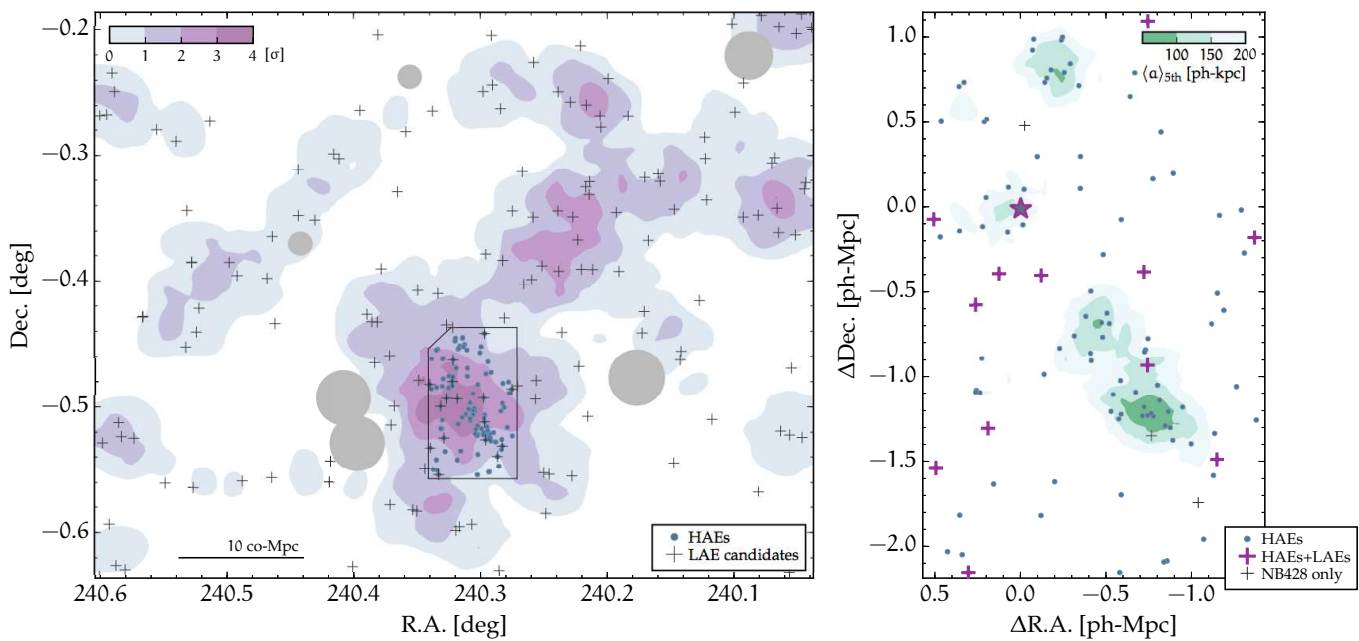}
    \caption{The 2-D maps of USS~1558 protocluster with Suprime-Cam (a: left) and 
    with MOIRCS (b: right). (a) The black crosses represent the LAE candidates, 
    and the blue circles indicate the HAEs identified by \citet{Hayashi:2016}. The 
    filled contours indicate the significance of LAE overdensities (0--$\sigma$, 
    $\sigma$--$2\sigma$, $2\sigma$--$3\sigma$, and $3\sigma$--$4\sigma$), which 
    are smoothed by the Gaussian kernel of $\sigma=1$ degree. The region enclosed 
    by the black lines corresponds to the survey field of MOIRCS for HAEs. (b) The 
    symbols are the same as shown in the left panel, but the purple crosses show 
    the dual \ha\ and \lya\ emitters. The star symbol indicates the RG. The filled 
    contours shows the mean distance of, 200--150, 150--100, and $<$100 ph-kpc 
    (physical kpc), smoothed by the Gaussian kernel of $\sigma=0.5$ arcmin.}
    \label{fig2}
\end{figure*}

Figure~\ref{fig2}a shows the spatial distribution of the LAE candidates over the 
entire field of Suprime-Cam. The protocluster cores traced by the HAEs with MOIRCS 
in our previous studies \citep{Hayashi:2012,Hayashi:2016} are embedded in the much 
larger-scale structures traced by LAEs. \citet{Hayashi:2016} have identified 100 
HAEs based on the combined technique of the narrowband selection and two 
colour--colour diagrams (r'JKs and r'H$_\mathrm{F160W}$Ks). These samples are 
limited to the star-formation rates (SFRs) of $>$2.2 \msun/yr without dust 
correction. 41 among those HAEs have been spectroscopically confirmed by 
\citet{Shimakawa:2014,Shimakawa:2015b}. Also from those spectroscopic analyses, we 
reckon that the contamination in the rest of our unconfirmed HAEs is less than 10 
\%. This protocluster core contains four very dense groups of HAEs; one in the 
immediate vicinity of the radio galaxy (RG), two toward the southwest (the further 
one is the densest), and one to the north of RG (Fig.~\ref{fig2}b). Within the 
MOIRCS survey field, we identify significant \lya\ emission lines for nine HAEs 
including the RG, which meet our LAE criteria. We also find four more objects 
which have both \lya\ and \ha\ detections in the two narrowbands, which were 
deselected from our original HAE samples in \citet{Hayashi:2016} by their 
colour--colour criteria. Including those, 104 HAEs are now identified in total as 
protocluster members, and 13 out of those are also classified as LAEs. On the 
other hand, three LAE candidates show no \ha\ emission line in our previous 
narrowband imaging at NIR.  Some of these are likely to be foreground 
contaminations (other emitters than \lya) and the rest would be too faint HAEs.

LAE density map shows a $4\sigma$ excess peak just around the mainbody of the 
protocluster USS~1558 traced by HAEs. The notable LSS or a gigantic filament 
extends toward northwest. A follow-up spectroscopy is needed to confirm the 
structures since our LAE samples would contain foreground other line 
contaminations. 
Surprisingly, however, on a much smaller scale ($\sim$300 physical kpc (ph-kpc)), 
HAEs with \lya\ line detections are distributed as if they are trying to avoid the 
overdense groups of HAEs (Fig.~\ref{fig2}b). It is remarkable that there is no LAE 
except for the RG in any of the notable dense group cores of HAEs in spite of the 
$4\sigma$ overdensity in LAEs in this protocluster as a whole on a $\sim$10 co-Mpc 
scale. To evaluate the deficiency of LAEs in the local overdensities, this work 
defines a density parameter, the mean projected distance 
$\langle{a}\rangle_\mathrm{Nth} = 2\times({\pi\Sigma_\mathrm{Nth}})^{-0.5}$
where $\Sigma_\mathrm{Nth}$ (=\ N/($\pi r_\mathrm{Nth}^2$)) is the number density 
of HAEs within the radius $r_\mathrm{Nth}$ which is the distance to the (N$-$1)th 
neighbours from each HAE. We use N=5. We stress that this density parameter 
maintains a relative consistency even if we choose different N values. 

\begin{figure}
	\centering
	\includegraphics[width=0.95\columnwidth]{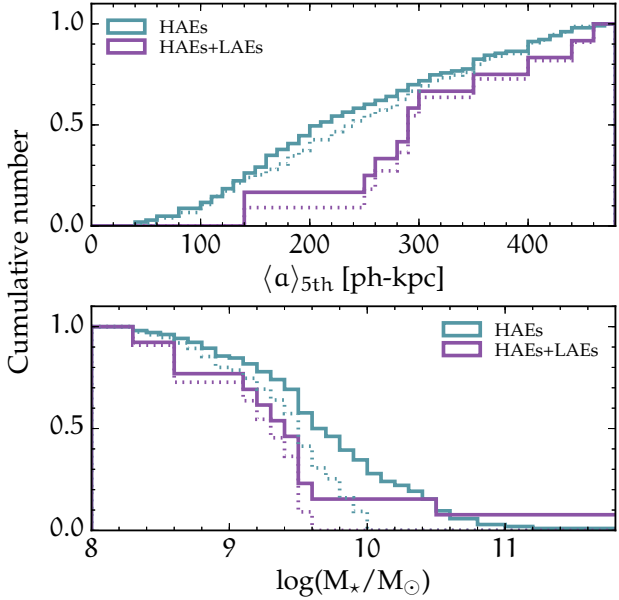}
    \caption{Normalised cumulative distributions of (a: upper) the mean distances 
    and (b: lower) stellar mass. The blue and purple solid lines indicate the 
    entire HAE samples and HAEs with significant \lya\ emission, respectively. The 
    dotted lines show the mass-control sample with M$_\star<$ 1E10$^{10}$ \msun.} 
    \label{fig3}
\end{figure}

The median value and the scatter of the mean projected distance 
($\langle{a}\rangle_\mathrm{5th}$) is $214_{-88}^{+140}$ ph-kpc. We investigate 
the significance of the LAE deficiency in the dense groups by dividing the HAE 
samples into high density and low density sub-samples separated at this median 
value. The fractions of LAEs among HAEs are $21\pm11$ \% in the lower densities 
and only $2\pm4$ \% excluding the RG (or $4\pm5$ \% with the RG) in the 
high-density regions, respectively. Fig.~\ref{fig3} represents the cumulative 
distributions of $\langle{a}\rangle_\mathrm{5th}$ (upper panel) and stellar mass 
(lower panel) for the HAEs and the HAEs with \lya\ emission detections. Stellar 
masses of the HAE samples are provided by \citet{Hayashi:2016}. In the upper 
panel, the possibility that ``HAEs" and ``HAEs+LAEs" are drawn from the same 
distribution is only 2 \% according to the Kolmogorov-Smirnov test, suggesting 
that \lya\ photons are more depleted in high-density regions. This trend is still 
statistically significant ($p$=0.04) even if we use the mass-control samples where 
we limit the galaxies only with the stellar masses lower than $10^{10}$ \msun. 
Even if we go further down in stellar masses, where such a statistical test would 
become no longer significant, the LAE deficiency at 
$\langle{a}\rangle_\mathrm{5th}<300$ ph-kpc would still remain. In fact, we do not 
see a significant difference in stellar mass distributions between the HAEs and 
those with \lya\ emission lines; $p$-value is 0.07 for the entire sample and is 
0.18 for the mass-control sample, respectively. Such an insignificant or small 
difference in stellar mass distributions between LAEs and non-LAEs is consistent 
with the recent studies \citep{Hagen:2016,Hathi:2016}. We stress that we here 
employ only the LAEs whose \ha\ emission lines are detected (as HAEs) by the 
independent narrowband imaging at NIR. Therefore, these comparisons are free from 
the contaminations in our LAE samples which could be a problem only for the 
LAE-only samples.

\section{Discussion and summary}

Our dual \lya\ and \ha\ line survey of a protocluster at $z=2.5$ presented has 
provided us with the first critical insight into the environmental dependence of 
the \lya\ strength as compared to \ha. The broad agreement in the spatial 
distributions between LAEs and HAEs on a large scale ($\gtrsim10$ co-moving Mpc) 
indicates that \lya\ line would be a good tracer of LSSs in the high-$z$ Universe. 
On a smaller scale, however, we see that LAEs, except for the RG, completely avoid 
the protocluster's dense cores, which are traced by the overdensities of HAEs. 
This means that the LAE surveys of protoclusters would inevitably miss the 
particularly dense regions of protoclusters which are likely to be the most 
interesting and critical environments where we expect to see any early 
environmental effects as the progenitors of present-day rich cluster cores. In 
other words, we are not really able to study environmental effects with the LAEs 
alone as they can trace only the outskirts of protocluster cores or even 
larger-scale structures around them. In order to search for truly dense structures 
at $z>2.6$ where \ha\ is no longer available, the James Webb Space Telescope 
\citep{Gardner:2006} will be very powerful as it probes the rest-frame optical 
regime where many nebular emission lines other than \lya\ are located. It should 
be noted, however, that bright \lya\ blobs can be also used as a tracer of the 
central galaxies in massive haloes as suggested by the past studies (e.g. 
\citealt{Steidel:2000,Matsuda:2011}). 

We measure the escape fraction of \lya\ photons by comparing the \lya\ and \ha\ 
fluxes. This can quantify the deficiency of LAEs in the protocluster's dense cores 
and thus provide insight into its physical origins. Unfortunately, the current 
datasets are not deep enough to estimate the escape fraction of \lya\ photons for 
individual HAEs, and thus we conduct the stacking analysis and derive \ha\ and 
\lya\ luminosities with high precision. The entire HAE samples are divided into 
two sub-samples by their local 2-D densities at the median value of the fifth mean 
distance (214 ph-kpc). The narrowband images are then combined with the 
{\sc imcombine} task by median on {\sc iraf}\footnotemark[2]. We derive \ha\ and 
\lya\ fluxes in the same way we measure the individual HAEs and LAEs in 
\citet{Hayashi:2016,Shimakawa:2016}. Photometric errors are estimated with a 
similar approach taken by \citet{Skelton:2014} where the 1-sigma Gaussian noise in 
background counts is measured as a function of variable aperture size. Our error 
measurements are thus performed independently of the SExtractor photometries since 
the SExtractor does not consider the pixel-to-pixel correlation and thus 
underestimates the errors especially for photometries with large aperture sizes. 
More details of the stacking method and the measurements of fluxes and errors will 
be presented in a forthcoming full paper \citep{Shimakawa:2017b}. We assume a 10 
\% flux contamination from \nii\ line to the narrowband flux for \ha, and 0.7 mag 
of dust extinction in \ha\ flux. The \lya\ escape fraction is given by 
$f_\mathrm{esc}^\mathrm{Ly\alpha} = (f_\mathrm{Ly\alpha,obs})/(8.7 f_\mathrm{H\alpha,int})$ 
where 8.7 is the ratio of \lya\ to \ha\ under the assumption of case B 
recombination \citep{Brocklehurst:1971}. This work estimates \lya\ and \ha\ fluxes 
with various aperture radii from 6 to 30 ph-kpc taking into account the fact that 
most of the star-forming galaxies show diffuse \lya\ components 
\citep{Ostlin:2009,Steidel:2011,Hayes:2013b}. 

\footnotetext[2]{\url{http://iraf.noao.edu}}

\begin{figure}
	\centering
	\includegraphics[width=0.95\columnwidth]{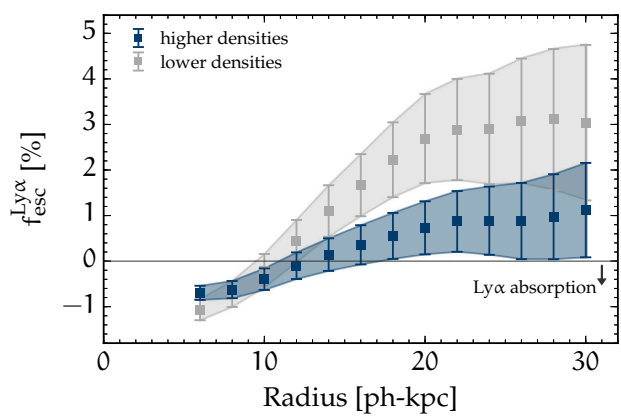}
    \caption{Based on the stacking analysis, the \lya\ photon escape fractions 
    ($f_\mathrm{esc}^\mathrm{Ly\alpha}$) of the composite HAEs are shown as a 
    function of photometric aperture radius within which we integrate line fluxes. 
    Here, the HAE samples are split into two sub-samples at the median value of 
    the mean projected distance ($\langle{a}\rangle_\mathrm{5th}=214$ ph-kpc). The 
    composite HAEs in the higher/lower density regions are presented by blue/grey 
    zones, respectively. The errorbars represent the photometric errors in \lya\ 
    and \ha\ line fluxes. The negative $f_\mathrm{esc}^\mathrm{Ly\alpha}$ values 
    mean that \lya\ is seen as an absorption line.} 
    \label{fig4}
\end{figure}

We compare the measured \lya\ photon escape fractions between the HAEs in 
high-density regions and those in lower-density regions. Both composite line 
images seem to have diffuse \lya\ profiles since the escape fraction increases 
with aperture radius, although photometric errors are quite large. In addition, we 
see \lya\ absorption features within the small aperture radii, which are 
consistent with the individual detection of \lya\ absorption in massive HAEs in 
the random field \citep{Shimakawa:2016}. Most importantly, we find systematically 
lower \lya\ photon escape fractions for the denser regions. This means that the 
\lya\ emission lines are systematically more depleted in denser regions than 
in lower-density regions in the protocluster (Fig.~\ref{fig4}). However, 
we should note that the \lya\ photon escape fraction is considered to depend on 
various physical properties such as dust, SFR, and metallicity 
\citep{Hayes:2010,Matthee:2016} that could depend on the environment. 
The discrepancy between the two composite HAEs may also contain such secondary
factors on top of the environment. Moreover, the narrowband technique cannot resolve 
the detailed spectral features around the \lya\ line. Thus our measurements may 
underestimate the escape fraction due to absorption by foreground 
circumgalactic/intergalactic medium (CGM/IGM) \citep{Hayes:2006} 
and/or blue-shifted dense outflowing gas \citep{Reddy:2016b}. 

In summary, we find that LAEs are missing in the dense HAE cores. We also find 
that \lya\ photon escape fractions in the composite HAEs in the denser regions 
are lower than those in lower-density regions. In the general field, it is 
expected that LAEs have less dust and lower \hi\ covering fractions
\citep{Shibuya:2014b,Reddy:2016b} and therefore \lya\ photons can relatively 
easily escape from the galaxies. However, in the protocluster core, 
we have extra surrounding gas and dust components trapped in the group/cluster 
scale haloes, which may prevent the \lya\ photons from escaping from the systems. 
Such abundant group-scale \hi\ gas may be supplied from the surrounding regions, 
for example, by cold streams \citep{Dekel:2009,Dekel:2009b}. Moreover, since the 
mean projected distance of 
HAEs is much smaller ($\lesssim200$ ph-kpc) in higher-density regions than in the 
lower-density regions, \lya\ photons emitted from a HAE have a higher chance of 
penetrating CGM associated to other member galaxy(ies) in the foreground located 
along the line of sight, and thus they may be more depleted. Otherwise, most of 
the individual galaxies in the dense cores could originally have lower \lya\ 
photon escape fractions due for example to higher dust extinction by a certain
environmental effect \citep{Koyama:2013b}. The current datasets 
are clearly insufficient to conclusively choose the most plausible explanation 
(further discussion will be described in \citealt{Shimakawa:2017b}).
A deep, rest-frame FUV spectroscopy of the protocluster galaxies with LRIS and/or 
KCWI on Keck telescope will be helpful for us to constrain the physical origins of 
the \lya\ depletion effect and to resolve the interplay between protocluster 
galaxies and CGM/IGM. 

\section*{Acknowledgements}

The data are collected at the Subaru Telescope, which is operated by the National 
Astronomical Observatory of Japan. This work is subsidized by JSPS KAKENHI Grant 
Number 15J04923. This work was also partially supported by the Research Fund for 
Students (2013) of the Department of Astronomical Science, SOKENDAI. We thank the 
anonymous referee for useful comments. R.S. and T.S. acknowledge the support from 
the Japan Society for the Promotion of Science (JSPS) through JSPS research 
fellowships for young scientists. T.K. acknowledges KAKENHI No. 21340045.




\bibliographystyle{mnras}
\bibliography{bibtex_library} 


\bsp	
\label{lastpage}
\end{document}